\documentclass{article}

\usepackage{arxiv}

\usepackage[utf8]{inputenc} % allow utf-8 input
\usepackage[T1]{fontenc}    % use 8-bit T1 fonts
\usepackage{hyperref}       % hyperlinks
\usepackage{url}            % simple URL typesetting
\usepackage{booktabs}       % professional-quality tables
\usepackage{amsfonts}       % blackboard math symbols
\usepackage{nicefrac}       % compact symbols for 1/2, etc.
\usepackage{microtype}      % microtypography
\usepackage{graphicx}
\usepackage{natbib}
\usepackage{doi}
\usepackage{here}
\usepackage{amsmath,amssymb}

\title{Bayesian inference is facilitated by modular neural networks with different time scales}

\author{ {\hspace{1mm}Kohei Ichikawa}\\
	Graduate  School  of  Arts  and  Sciences\\
	The  University  of  Tokyo\\
	Meguro-ku, Tokyo 153-8902, Japan \\
	\And
	{\hspace{1mm}Kunihiko Kaneko} \\
	The Niels Bohr Institute\\
	University of Copenhagen\\
	Blegdamsvej 17, Copenhagen, 2100-DK, Denmark \\
	\\
}

\renewcommand{\shorttitle}{Bayesian inference in neural networks with different time scales}

\hypersetup{
pdftitle={ Bayesian inference is facilitated by modular neural networks with different time scales},
pdfsubject={q-bio.NC},
pdfauthor={Kohei Ichikawa, Kunihiko Kaneko},
pdfkeywords={First keyword, Second keyword, More},
}

\begin{document}
\maketitle

\begin{abstract}
Various animals, including humans, have been suggested to perform Bayesian inferences to handle noisy, time-varying external information.
In performing Bayesian inference, the prior distribution must be shaped by sampling noisy external inputs. 
However, the mechanism by which neural activities represent such distributions has not yet been elucidated. 
In this study, we demonstrated that the neural networks with modular structures including fast and slow modules effectively represented the prior distribution in performing accurate Bayesian inferences. 
Using a recurrent neural network consisting of a main module connected with input and output layers and a sub-module connected only with the main module and having slower neural activity, we demonstrated that the modular network with distinct time scales performed more accurate Bayesian inference compared with the neural networks with uniform time scales. 
Prior information was represented selectively by the slow sub-module, which could integrate observed signals over an appropriate period and represent input means and variances. 
Accordingly, the network could effectively predict the time-varying inputs. 
Furthermore, by training the time scales of neurons starting from networks with uniform time scales and without modular structure, the above slow-fast modular network structure spontaneously emerged as a result of learning wherein prior information was selectively represented in the slower sub-module. 
These results explain how the prior distribution for Bayesian inference is represented in the brain, provide insight into the relevance of modular structure with time scale hierarchy to information processing, and elucidate the significance of brain areas with slower time scales.
\end{abstract}

% keywords can be removed

\section{Introduction}
In the human and various animal brain, information processing involves inference based on inputs from the external world through the sensory systems, which obtains information with uncertainty due to noise. 
Previous studies suggested that animals such as humans and monkeys process inputs according to a Bayesian inference framework to deal with such uncertainty\citep{KNILL2004712, ANGELAKI2009452, HAEFNER2016649, Ernst2002, FRISTON20121230, Merfeld1999, Doya2007-yy, probabilistic_brain, Beck15310, illusion_perception, doi:10.1073/pnas.1918143117}.

Bayesian inference is performed by calculating the posterior from the prior, which refers to the information possessed in advance about the signal, and the likelihood estimated by observing the input signal. 
Hence, it is believed that the prior must first be represented in the brain, but how prior information is shaped in the brain remains unclear.
In previous studies, the prior has often been treated as a given value\citep{Echeveste:2020aa}, and the mechanism for shaping the prior by learning has not been considered. 
Evolutionary acquisition of the prior has been proposed\citep{universal_darwinism, evolution_bayesian}, whereas it is naturally expected that such information should be shaped within one generation through observing and learning time-dependent signals.
Experimental results suggest that the prior and the likelihood for Bayesian inference are encoded in different brain areas\citep{differential_representation, Chan7817, dAcremont10887}. 
Still, the validity and the mechanisms underlying the results remain controversial, and how area differentiation is relevant to the accuracy of Bayesian inference is not well understood.
A simulation\citep{population_codes_of_prior_knowledge} suggested that a gain of the activation function encodes the prior.
However, because the prior was fixed in this study, how shaping occurs when the prior varies over time was not considered.

In general, to obtain the prior, it is necessary to estimate the prior distribution based on previous observations, and the population of neurons that represents the prior must integrate observed inputs over time.
One possible mechanism for achieving such integration may be two neural modules functioning at distinct time scales: a downstream neuron population with slower activity changes separated from an upstream neuron population that processes input information.
In this structure, the slow module that does not directly receive inputs may facilitate integration. 
Some experimental reports have suggested that the time scale of neural activities in downstream areas of the brain that do not directly receive external input is slow\citep{hierarchy_of_intrinsic_timescale, diversity_intrinsic_timescale, brain_and_its_time}. 
On this basis, we evaluated recurrent neural networks (RNNs) with two modules; a main module with direct connection to the input-output layer and a sub-module with a direct connection to the main module and no connections to the input-output layer (i.e., a hierarchical structure)(Fig.\ref{fig:hierarchical_rnn}). 
Then, we examined the role of modular structure and the relevance of the time scale difference between the main and sub-modules for the prior representation for Bayesian inference.

We found that RNNs with a modular structure shape the prior more appropriately than regular RNNs. 
Further, Bayesian inference is more accurate when the time scale of the sub-module is appropriately slow. 
When the time scale is uniform, prior information is maintained in both the main module and sub-module. 
In contrast, when the time scales are different, prior information is represented by the slow sub-module. 
Comparing these two cases revealed that the coded variance of prior on the neural manifold was easier to decode in the time scale difference model, which facilitated the distinction of the average input change from noise. 

In addition, we examined if the modular structure with distinct time scales would emerge from a homogeneous neural network.
We trained the network in a Bayesian inference task where the time scale of each neuron varied in time. 
As the training progressed, we observed that the time scales of neurons differentiated into slower and faster scales.
A modular structure arose in which slow neurons were separated from the input/output layers, which were predominantly connected to the fast neurons, and a sub-module with slow neurons represented the prior information.

These results are crucial for understanding the prior representation mechanism in Bayesian inference and provide insight into the relationships between neural network structure, neural dynamics\citep{AmuntsENEURO.0316-21.2022, MASTROGIUSEPPE2018609, Computation_Through_Neural_Population_Dynamics, Shaping_dynamics}, and time scales\citep{timescale_in_cognitive_neuroscience} underlying information processing in the brain, which is considered the central issue of computational neuroscience.

\section{Model}
\subsection{Recurrent Neural Networks with/without modular structure}

To investigate the effect of structure and time scale on Bayesian inference, we considered the following RNNs\citep{versatile}.

\begin{figure}
  \centering
  \includegraphics[width=12cm]{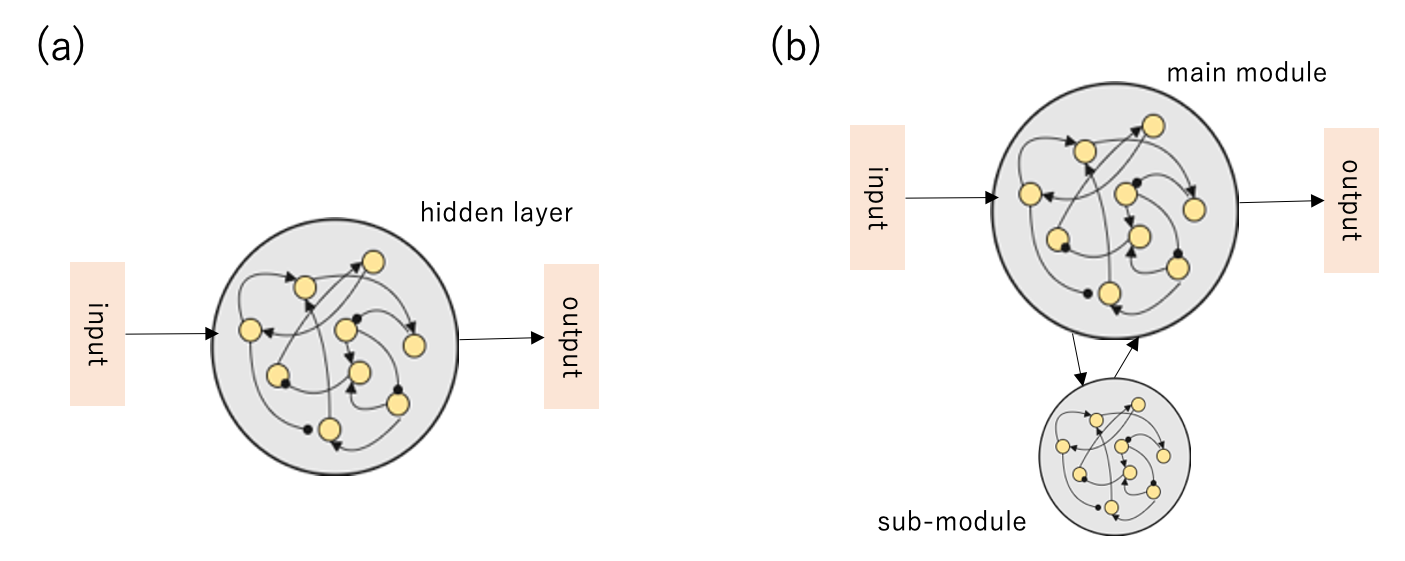}
  \caption{Schematic of RNN. (a) Standard RNN without modular structure (b) RNN with modular structure}
  \label{fig:hierarchical_rnn}
\end{figure}

First, we established a regular RNN consisting of an input layer, a recurrent(hidden) layer, and an output layer, as shown in Fig.\ref{fig:hierarchical_rnn}(a).
The following equation represents the dynamics of the recurrent layer:

\begin{equation}
    {\bf x}(t+1) = ({\bf I}-{\boldsymbol \alpha}){\bf x}(t) + {\boldsymbol \alpha} {\rm ReLU}(W_{in}{\bf u}(t)+W{\bf x}(t)) + \sqrt{\boldsymbol \alpha}{\boldsymbol \xi},
    \label{eq:regular_network_dynamics}
\end{equation}
where ${\boldsymbol \alpha} = (\alpha_1, \alpha_2, ..., \alpha_{200})^\mathsf{T}$ represents a vector to introduce the time scale of the neurons as 

\begin{equation}
    \alpha_i  =  \left\{  \begin{array}{ll}
        \alpha_m  & (1 \leq i \leq 150)\\
        \alpha_s  &  (150 \leq i \leq 200), \\
        \end{array}
	\right.
    \label{eq:alpha_separation}
\end{equation}
where the standard homogeneous network is given by $\alpha_s=\alpha_m$; the case with $\alpha_s < \alpha_m$ was also studied to investigate the effect of time scale difference.
Although we mainly studied the systems with 150 fast, and 50 slow neurons, the results to be discussed are not altered, as long as both the numbers are sufficient (say  100 vs 50, 150 vs 150 for fast and slow neurons).
Here, ${\bf u}(t)$ is the input signal, and ${\bf x}$ is the state of the neurons in the recurrent layer. 
We adopted the activation function ReLU(${\rm RELU}(z)=0$ for $z \leq 0$ and $=0$ for $z > 0$)\citep{ReLU}. 
Then, the output of the RNN was determined by the linear combination of the internal states as follows.

\begin{equation}
    {\bf y}(t) = W_{out}{\bf x}(t)
    \label{output_calculation}
\end{equation}
In Eq.\ref{eq:regular_network_dynamics}, ${\boldsymbol \xi}$ was used to account for noise in dynamics given by a random variable that follows a normal distribution with mean $0$ and standard deviation $0.05$.

Next, we introduced a modular structure to the above RNN to ensure the distinction of main and sub-modules(Fig.\ref{fig:hierarchical_rnn}(b)). 
Only the main module was connected to the input/output layers. Thus, the dynamics of the recurrent layer are given by 

\begin{equation}
        {\bf x}_m(t+1) = (1-\alpha_m){\bf x}_m(t) + \alpha_m {\rm ReLU}(W_{in}{\bf u}(t)+W_{main}{\bf x}_m(t)+W_{s\rightarrow m}{\bf x}_s(t)) + \sqrt{\alpha_m}{\boldsymbol \xi}_m
    \label{eq:modular_network_main}
\end{equation}
\begin{equation}
        {\bf x}_s(t+1) = (1-\alpha_s){\bf x}_s(t) + \alpha_s {\rm ReLU}(W_{sub}{\bf x}_s(t)+W_{m\rightarrow s}{\bf x}_m(t)) + \sqrt{\alpha_s}{\boldsymbol \xi}_s,
    \label{eq:modular_network_sub}
\end{equation}
where ${\bf x}_m and {\bf x}_s$ represent the firing rate of neurons in the main and sub-modules, respectively. 
Here, $\alpha_m and \alpha_s$ represent the time scale of the main and the sub-module, respectively. $\alpha_m$ is fixed at $1$, while we varied $\alpha_s$ from $1$ to $0.01$ to examine the effect of the time scale difference. 
The RNN output was determined by the linear combination of internal states of the main module.

\begin{equation}
    {\bf y}(t) = W_{out}{\bf x}_m(t)
    \label{eq:output_calculation_modular}
\end{equation}

\subsection{Task}
In this study, we considered a task in which Bayesian inference improves estimation accuracy. Specifically, the RNN was tasked with estimating the true value from an observed signal with noise. 
We generated the external input as follows:
First, the true value $y_{true}$ was randomly sampled from a generator(cause) distribution, given by the normal distribution with mean $\mu_g$ and variance $\sigma_g^2$. 
Next, the observed signal $s$ was generated from $y_{true}$ by adding noise so that the input is given by the normal distribution with mean $y_{true}$ and variance $\sigma_s^2$.
The generator did not remain constant: It changed with probability $p_t$ over time. 
When the generator changed, $\mu_g, \sigma_g$ were sampled uniformly from $\mu_g \in [-0.5, 0.5], \sigma_g \in [0, 0.8]$ respectively.

As mentioned in the Introduction, the prior distribution needed for Bayesian estimation must be estimated from the observed signal so that it is close to the generator distribution.
Then, ${\bf u}(t)$ for Eq.\ref{eq:regular_network_dynamics} (or \ref{eq:modular_network_main},\ref{eq:modular_network_sub}) is given by using the Probabilistic Population Code (PPC), which has been proposed as the neural basis for Bayesian inference\citep{Ma2006}.
PPC assumes that the information in a signal is encoded by a population of neurons with a position-based preferred stimulus that fires probabilistically according to a Poisson distribution. 
It has been shown that neural networks with a population of neurons following PPC as the input layer can learn probabilistic inference effectively\citep{Orhan2017}. 
Therefore, in this study, we also assumed that the activity ${\bf u}$ of the input-layer neurons encoding the observed signal followed the PPC model. 
${\bf u}$ was sampled from the following Poisson distribution\citep{PPC_sampling}:

\begin{equation}
    p({\bf u}|s) = \prod_i \frac{e^{-f_i(s)}f_i(s)^{u_i}}{u_i!}
    \label{eq:poisson}
\end{equation}

Here, $s$ is the observed signal generated from $y_{true}$ by adding noise, and $f_i$ is the tuning curve of the neurons. 
This selective firing occurs in proportion to the gain when the observed signal is generated.
This gain is inversely proportional to the noise variance as $g=1/\sigma_l^2$, and corresponds to signal clarity. 
Namely, the gain  decreases and noise increases due to uncertainty in observations\citep{TOLHURST1983775}.
Considering the gain, we obtain:

\begin{equation}
    f_i(s)=g\exp \biggl(\frac{-(s-\phi_i)^2}{2\sigma_{\rm PPC}^2}\biggr),
    \label{eq:tuning_function}
\end{equation}
where $\phi_i$ represents the preferred stimuli of neurons in the input layer. 
It was assumed that $\phi_i$ follows an arithmetic sequence for $i$ ($\phi_i=-1/2+i/m$ when the number of neurons in the input layer is $m$)\citep{tuning_curve}. 
Also, $\sigma_{\rm PPC}^2$ is a constant that represents the ease of firing and was set as $\sigma_{\rm PPC}^2=1/2$ in this study.

In this task, the true value $y_{true}$ was to be estimated based on the input signal ${\bf u}$. 
Therefore, training was performed to minimize the mean squared error (MSE) between the neural network output $y(t)$ and the true value $y_{true}(t)$.  
Note that the loss function was not based on the Bayesian optimal value calculated from the generator distribution and the noise in the observed signal but only calculated based on the true value.

\begin{equation}
    L = \frac{1}{T}\sum_t (y(t)-y_{true}(t))^2
    \label{eq:loss_function}
\end{equation}

Training was performed by the backpropagation method \citep{backpropagation_original, bptt}. 
An efficient Stochastic Gradient Descent method, Adam\citep{adam}, was used for optimization.
The batch size of training samples was set to 50, and the weight decay rate was set to 0.0001; training was performed for 6000 iterations(See Table.\ref{table:hyperparameters} for the hyperparameters used in the experiment).

\begin{table}
    \caption{Hyperparameters}
	\begin{center}
            \begin{tabular}{c|c}  \hline\hline
                Attribute  &  Value  \\  \hline\hline
	        Range  of  $\mu_p$  &  $-0.5\leq  \mu_p \leq  0.5$    \\
                Range  of  $\sigma_p$  &  $0  \leq  \sigma_p \leq  0.8$  \\
                Range  of  $\sigma_l$  &  $\sqrt{1/5}  \leq  \sigma_l \leq  1$  \\
                Switching probability of prior & $p_t=0.03$ \\
                Length  of  ${\bf  u}(t)$  &  $100$  \\
                $\sigma_{\rm  PPC}$  &  0.5  \\
                Lasting  time  of  ${\bf  u(t)}$  &  $T=120$  \\
                \#Neurons  in  the  main module  &  150  \\
                \#Neurons  in  the  submodule  &  50  \\
                $\alpha_m$  &  1  \\
                $\alpha_s$  &  1, 0.5, 0.2, 0.1, 0.05, 0.01  \\
                Batch  size  &  50  \\
                Optimization  algorithm  &  Adam  \\
                Learning  rate  &  0.001  \\
                Iteration  &  6000  \\
                Weight  decay  &  0.0001  \\  \hline
                \end{tabular}
	\end{center}
    \label{table:hyperparameters}
\end{table}

\section*{Results1: Fixed structure and time scales}
\subsection*{Bayesian optimality}
Because the generated signal $s$ was observed under noise, the neural network was required to estimate the true value sampled from the generator. 
If the information from the generator was known, $y_{true}$ would be estimated by minimizing the long-term MSE, which reveals the optimal $y$ value as follows (maximum a posteriori(MAP) estimation \citep{PRML}).

\begin{equation}
    y_{opt}=\frac{\sigma_g^2}{\sigma_g^2+\sigma_s^2}s+\frac{\sigma_s^2}{\sigma_g^2+\sigma_g^2}\mu_g
    \label{eq:y_opt}
\end{equation}

However, as described in the “Task” section, the information from the generator was not explicitly given to the neural network, so it must be estimated from observed signals as a prior distribution.
First, we examined whether the neural network could achieve this prior-based estimation.

\begin{figure}
  \centering
  \includegraphics[width=10cm]{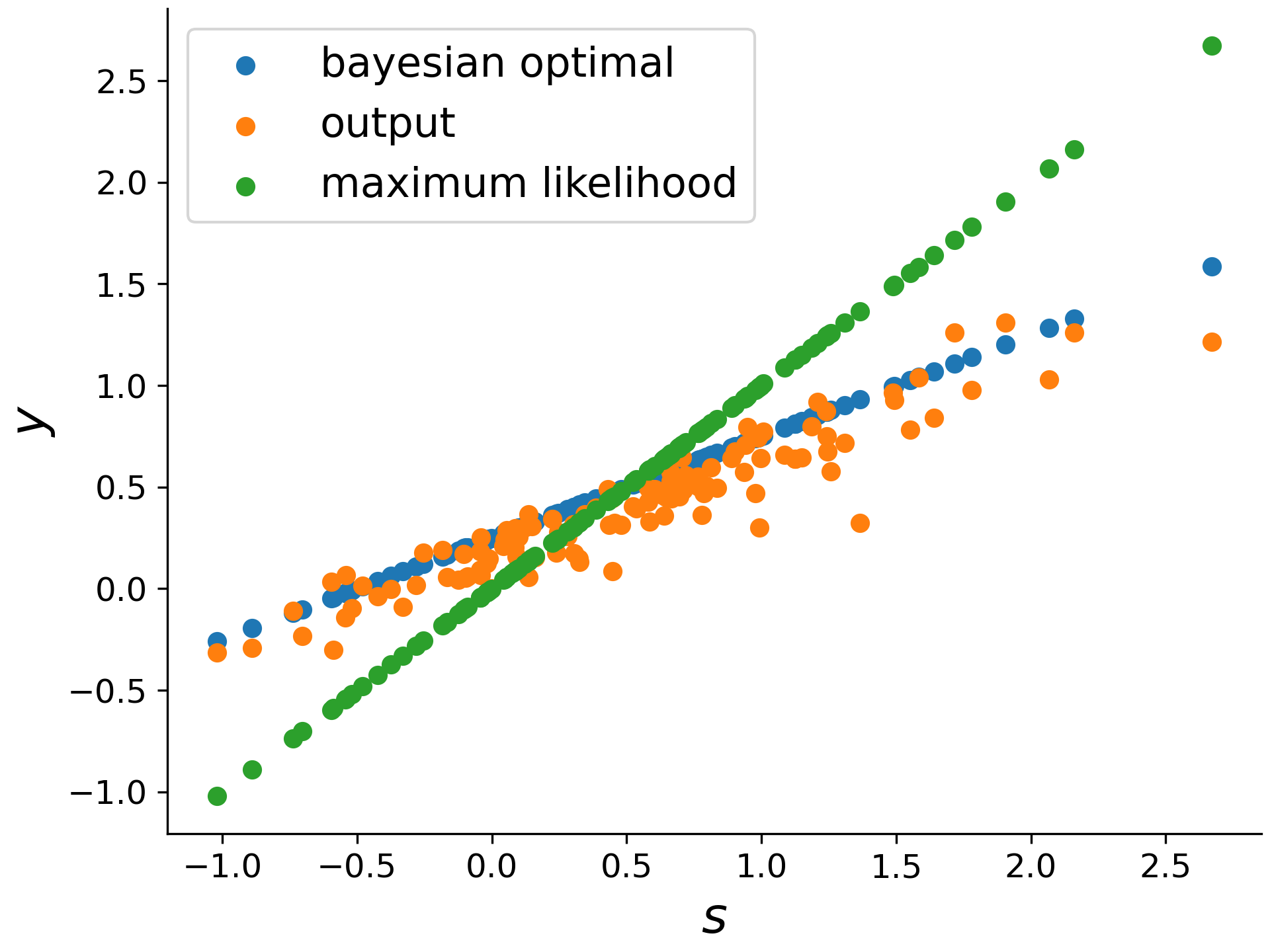}
  \caption{The output $y$ of RNN against the observed signal value $s$. 
  Before $s$ is input, the time series signal, which is sampled from the normal distribution with the mean $\mu_g=0.5$ and the standard deviation $\sigma_g=0.5$ and then the noise with the standard deviation $\sigma_s=\sqrt{1/5}$ is added in the input.
  The accuracy can be increased by estimating prior based on the signal input before $s$ and performing Bayesian inference.
  Blue points represent $y_{opt}=\frac{\sigma_g^2}{\sigma_g^2+\sigma_s^2}s+\frac{\sigma_s^2}{\sigma_g^2+\sigma_g^2}\mu_g$ value, orange points represent the output of RNN $y$, and green points represent estimation based on maximum likelihood estimation $y_{ML}=s$. The result is for a model with $\alpha_s=0.1$.}
  \label{fig:output}
\end{figure}

The output $y$ of RNN with modular structure trained with $\alpha_s=0.1$, when given an observed signal $s$, is shown Fig.\ref{fig:output}. 
$s$ was sampled from the prior with $\mu_g=0.5, \sigma_g=0.5$, and $\sigma_s=\sqrt{1/5}$ of noise was added.
The green points represent the estimation based on the maximum likelihood estimation $y_{ML}$, which is that with the highest accuracy when no prior information is available. 
Here, this estimation is equal to the observed signal $s$.
The blue points represent $y_{opt}$ when estimated according to the MAP estimation, and the orange points represent the actual neural network output $y$.
Fig.\ref{fig:output} shows that the output of RNN is closer to the blue points $y_{opt}$ rather than to the green points, indicating that approximate Bayesian inference (Near-optimal Bayesian inference) with a well-estimated prior is achieved (the mean squared error between $y$ and $y_{ML}$ is $0.15$, and the mean squared error between $y$ and $y_{opt}$ is $0.019$, the latter being smaller).

\begin{figure}
  \centering
  \includegraphics[width=15cm]{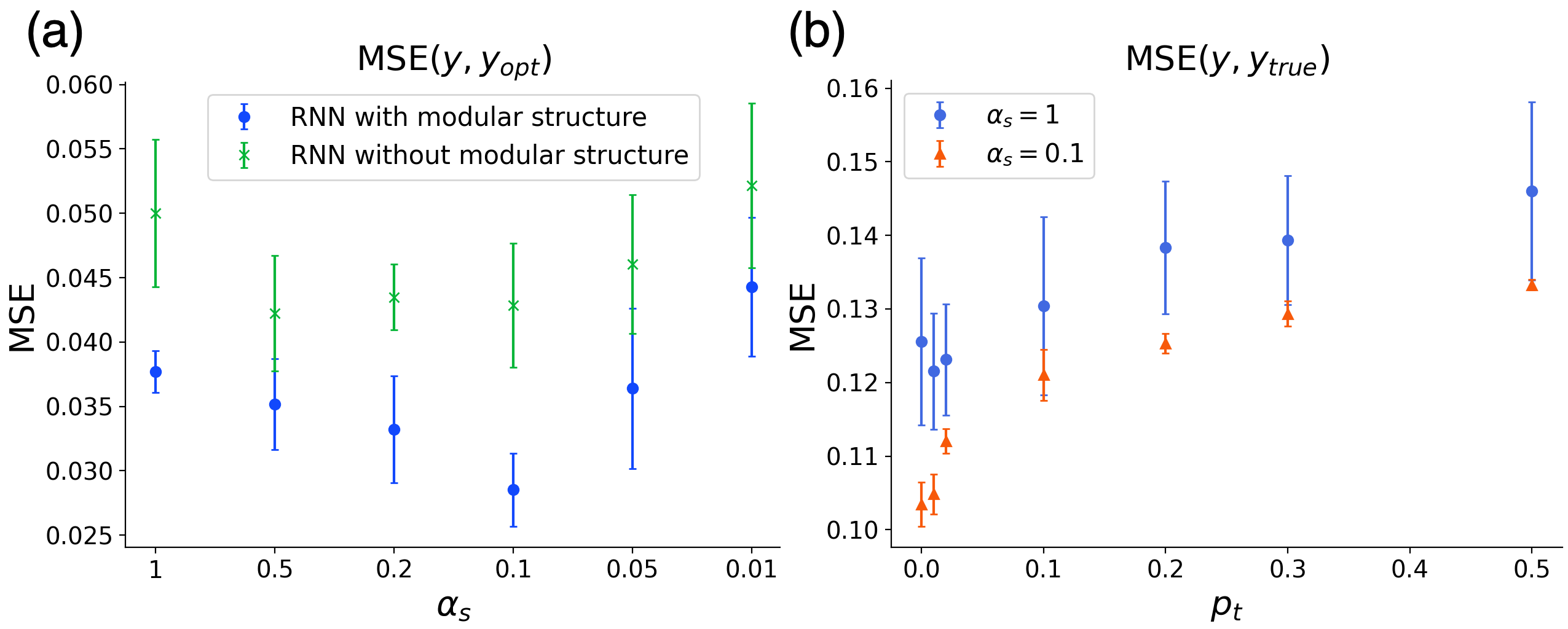}
  \caption{
    (a) MSE between the optimal value $y_{opt}(t)$ and the output of RNN $y(t)$, plotted against the time scale $\alpha_s$. $\cdot$ with modular and $\times$ without modular structure.
    RNNs with a modular structure is more accurate. 
    In addition, those with $\alpha_s \sim (0.1\sim 0.2)$ have optimal error.
    (b) MSE between the true value $y_{true}(t)$ and the output of RNN $y(t)$ for the network with $\alpha_s=1(\cdot)$ and $\alpha_s=0.1(\times)$. The value increases as $p_t$ increases, but the model with $\alpha_s=0.1$ is always more accurate.
  }
  \label{fig:bayesian_optimality}
\end{figure}

Next, we examined the optimality of the Bayesian estimation for networks with and without modular structures and time scale differences. 
Fig.\ref{fig:bayesian_optimality}(a) shows the MSE between $y$ and $y_{opt}$ by the RNN trained under each condition. 
This result shows that the modular structure improved the accuracy of Bayesian estimation, which was further increased when $\alpha_s$ decreased to an appropriate degree. 
In fact, we found the optimal time scale $\alpha_s=0.1\sim 0.2$, at which maximum accuracy was achieved.
Even without modular structure, the time scale difference contributed to inference accuracy, but the accuracy increased significantly with both the modular structure and time scale difference.

\subsection*{Adjustability to rapid generator switching}
So far, we studied the performance of Bayesian inference models under a fixed generator to compare the accuracy of Bayesian inference itself.
Next, we examined their performance when the generator changes in time.
To perform Bayesian inference for a rapidly changing input, it was necessary for the model to quickly approach the new optimal value $y_{opt}$ to yield a good estimation. 
To verify the accuracy of the RNN in this case, we compared the MSE between $y_{true}(t)$ generated by the generator and the output $y(t)$ of RNN under various $p_t$(Fig.\ref{fig:output}(b)). 
The model with $\alpha_s=0.1$ was found to be more accurate for all values of $p_t$.

As a special case, we considered a setting where the input moves back and forth between two generators, A and B. 
Then we examined whether the prior distribution estimated by the RNN was closer to the distribution of either generator.
Specifically, we adopted the generator A with $(\mu_g, \sigma_g^2)=(\mu_A, \sigma_A^2)$ and the generator B with $(\mu_g, \sigma_g^2)=(\mu_B, \sigma_B^2)$ and computed the following values when the Bayesian optimal estimates under each generators were $y_{opt}^A, y_{opt}^B$.
\begin{equation}
    a(t) = \frac{y_{opt}^B-y(t)}{y_{opt}^B-y_{opt}^A}
    \label{eq:adjustability}
\end{equation}
When $a(t)$ is close to 1, the model's prior is closer to generator B, and when $a(t)$ is close to -1, it is closer to prior A.

\begin{figure}
  \centering
  \includegraphics[width=15cm]{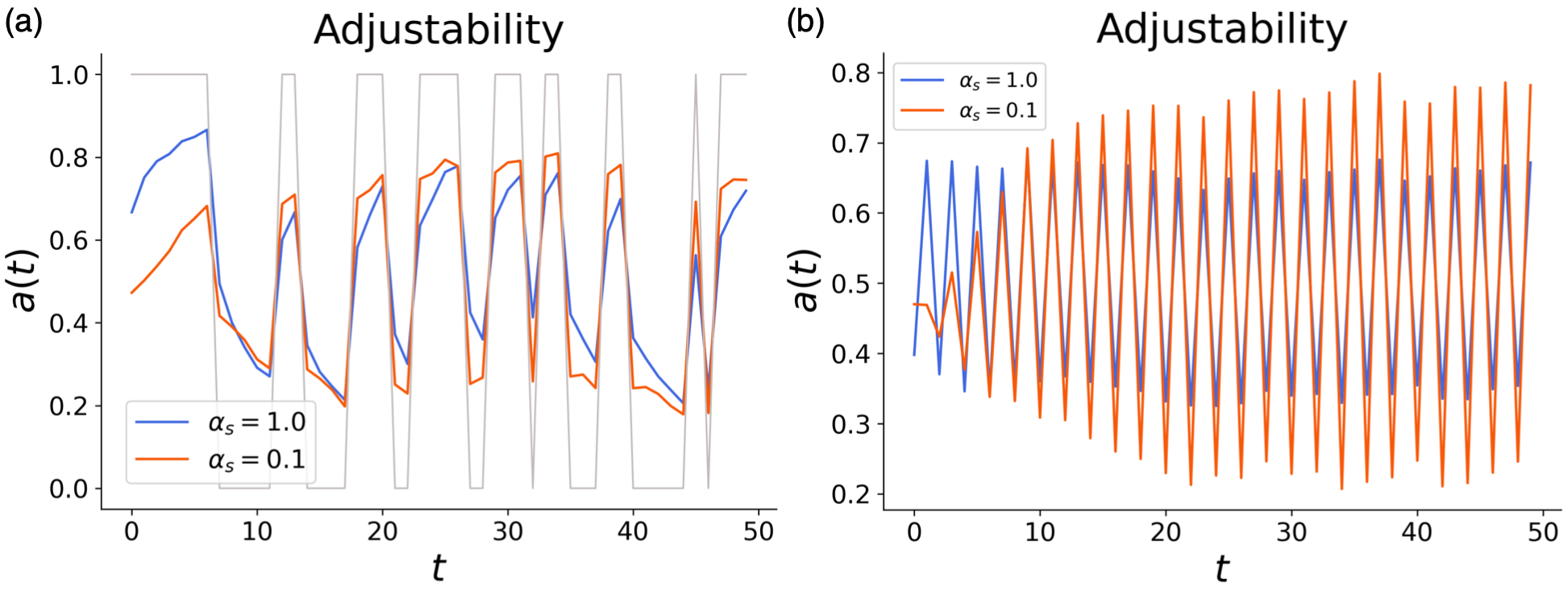}
  \caption{
    Adjustability to rapid generator change. 
    (a) $a(t)$ for the case where generator A($(\mu_A, \sigma_A^2)=(-0.5, 0.04)$) and generator B($(\mu_A, \sigma_A^2)=(0.5, 0.04)$) switch alternately with probability $p_t=0.2$. 
    The model with $\alpha_s=0.1$ adjusted more quickly to the generator change.
    The thin line represents $a(t)$ when the output was fully adjusted to generator switching.
    (b) $a(t)$ for the case where generator A($(\mu_A, \sigma_A^2)=(-0.5, 0.04)$) and generator B($(\mu_A, \sigma_A^2)=(0.5, 0.04)$) switch every time(periodic switch).}
  \label{fig:adjustability}
\end{figure}

Comparing the change in $a(t)$ between the model with $\alpha_s=0.1$ and the model with $\alpha_s=1$, we found that the model with $\alpha_s=0.1$ was more adjustable to the generator change as shown in Fig.\ref{fig:adjustability}(a). 
This result shows that the model with $\alpha_s=0.1$ was more responsive to the changes of the generators and recognized the generator change more quickly in all runs. 
The difference between the two models was especially pronounced in the extreme case in which the two generators switched every time(Fig.\ref{fig:adjustability}(b)). 
Intuitively, having a population of slow neurons would seem to be a disadvantage in responding to rapid environmental changes, but the results showed the opposite.  
The network with $\alpha_s=1$ could not follow rapid input changes, whereas that with $\alpha_s=0.1$ could estimate the input prior effectively.
We discuss the importance of slow neurons in responding to rapid changes below.

\subsection*{Representation of the prior}
We investigated how the slow sub-module facilitated improved prior representation for Bayesian inference. 
Beginning from the hypothesis that a group of downstream slow neurons represent the prior by integrating the observed signal over time, we investigated which side of the main/sub-module was responsible for the prior information in the modular RNN.

Here, by using the prior information, the estimated value was shifted from the observed signal $s$ to an appropriate value $y_{opt}$(Eq.\ref{eq:y_opt}).
In other words, even given the same signal input $s$, the output varied depending on which time series signal was input before $s$ (because the prior estimation changed). 
Even if one module returned to its original state, the output shifted from $s$ because the prior information remained in the other module. 
The scale of this change is considered to represent the degree to which the module utilizes the estimated prior information.
Therefore, it is possible to estimate the extent to which each module plays a role in prior information processing by examining the change in the output $y(t)$ when the internal state of each main and sub-modules is changed to the value corresponding to a different prior.

First, let ${\bf x}_m(\mu_g, \sigma_g), {\bf x}_s(\mu_g, \sigma_g)$ be the internal states of the main and sub-module, respectively when the input signal $s$ from a generator $(\mu_g, \sigma_g)$ is applied for a certain period. 
Because the output $y$ is determined by the internal states of two modules and the input signal, it can be written as $y({\bf x}_m(\mu_g, \sigma_g); {\bf x}_s(\mu_g, \sigma_g), s, \sigma_s)$.
From this, the change in output $y$ is computed by fixing one of the two modules and varying the other to a different internal state ${\bf x}_i(\mu_g, \sigma_g) \rightarrow {\bf x}_i(\mu_g', \sigma_g')$.
The degree of change in $y$ represents the impact on the output of each module reflecting the prior information.
Hence, by comparing the above variances of $y$ by ${\bf x}_m$(or ${\bf x}_s$) with fixed ${\bf x}_s$(or ${\bf x}_m$) respectively, it is possible to estimate how much each module is responsible for the prior representation.
Specifically, we fixed one of the modules at $\mu_g=0$, $\sigma_g=0.4$ (These values are set to the median of the range of values $-0.5 \leq \mu_g \leq 0.5$, $0\leq \sigma_p \leq 0.8$), i.e. ${\bf x}_i(0, 0.4)$, while for the other module $\mu_g$ and $\sigma_g$ are changed as ${\bf x}_g(\mu_g, \sigma_g)$.
Then, we calculated the variance of $y$ as

\begin{equation}
    V_m = \langle {\rm Var}[y({\bf x}_m(\mu_p, \sigma_p), {\bf x}_s(0, 0.4), s, \sigma_l)]_{(\mu_p, \sigma_p)}\rangle _{(s, \sigma_l)}
\end{equation}
\begin{equation}
    V_s = \langle {\rm Var}[y({\bf x}_m(0, 0.4), {\bf x}_s(\mu_p, \sigma_p), s, \sigma_l)]_{(\mu_p, \sigma_p)}\rangle _{(s, \sigma_l)},
\end{equation}
where ${\rm Var}[\ ]_{(\mu_p, \sigma_p)}$ denotes the variance over the changes of $(\mu_p, \sigma_p)$, and $\langle \rangle_{(s, \sigma_l)}$ denotes the average over the changes of $(s, \sigma_l)$.
The larger $V_s$ or $V_m$ indicates that the sub-module or main module strongly reflects the difference in the prior distribution to the difference in output, respectively.

\begin{figure}
  \centering
  \includegraphics[width=8cm]{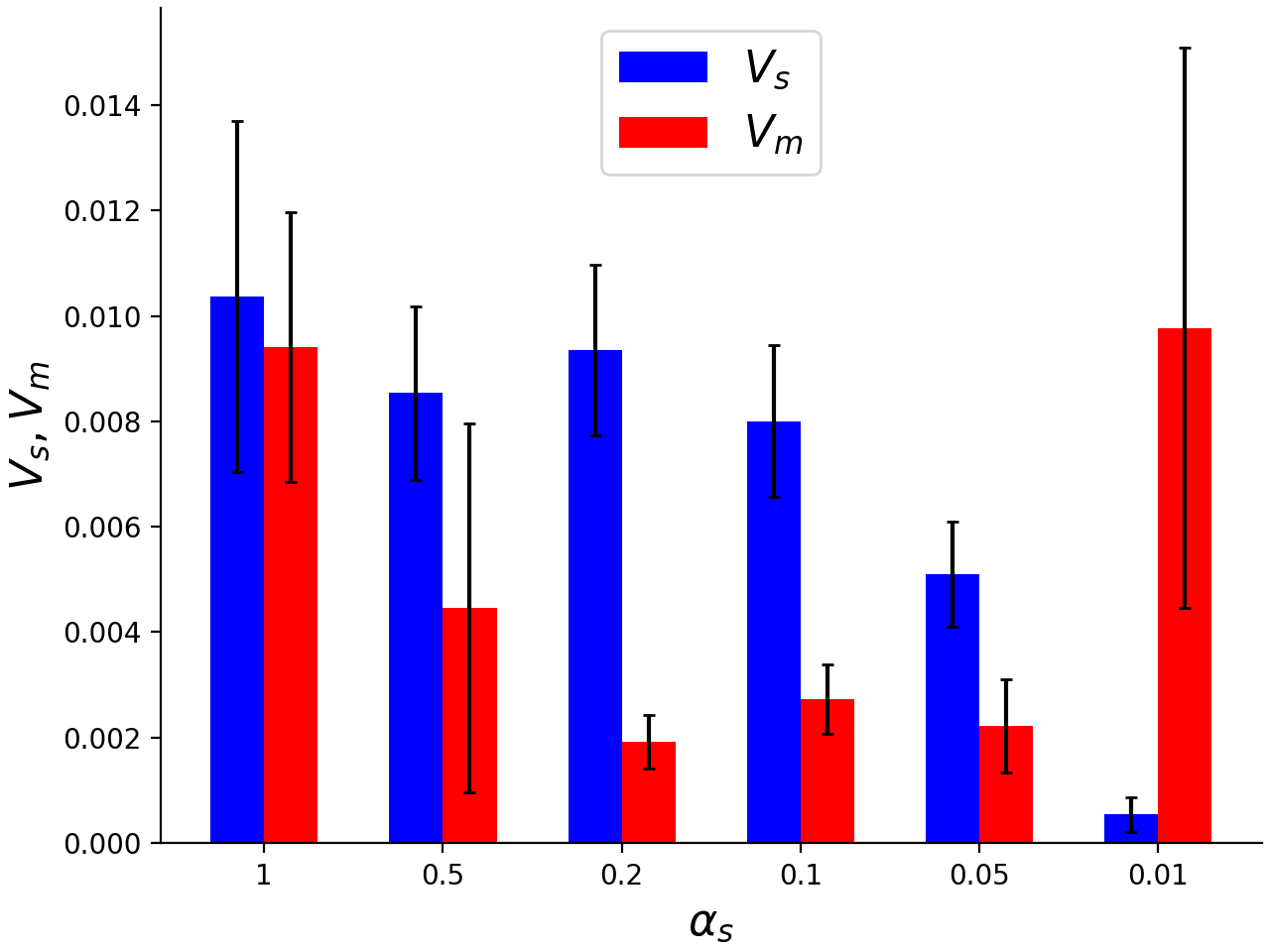}
  \caption{
    Division of roles for representing prior distribution. 
    $V_s, V_m$ defined in the text Eqs. (12,13) plotted for different values of $\alpha_s$ computed over $1000$ samples of data. 
    $V_s$ and $V_m$ represent the degree to which the sub-module and the main module are responsible for prior-based information processing. When $\alpha_s=0.2, 0.1$, the sub-module selectively contributes to the representation of the prior.
    }
  \label{fig:differentiation_of_prior_representation}
\end{figure}

Dependencies of $V_s$ and $V_m$ on different $\alpha_s$ are shown in Fig.\ref{fig:differentiation_of_prior_representation}.
This result shows that when $\alpha_s=1$ (i.e., the time scale is uniform), both the main and sub-modules contribute to the representation of prior distribution to the same degree. 
Conversely, when $\alpha_s=0.1\sim 0.5$, $V_s$ is much larger than $V_m$, meaning that the sub-module selectively contributes to the representation of the prior. In particular, when $\alpha_s=0.1$ and $0.2$, the differentiation of representation between the main and sub-modules is more pronounced. 
Note that the contribution of the main module is large when $\alpha_s=0.01$, probably because the time scale of the sub-module is too slow to code the information of the prior. 
Comparing of Fig.\ref{fig:differentiation_of_prior_representation} and Fig.\ref{fig:bayesian_optimality} shows that the highly accurate Bayesian inference is achieved when the prior distribution information is localized in the sub-module.

\begin{figure*}
  \centering
  \includegraphics[width=15cm]{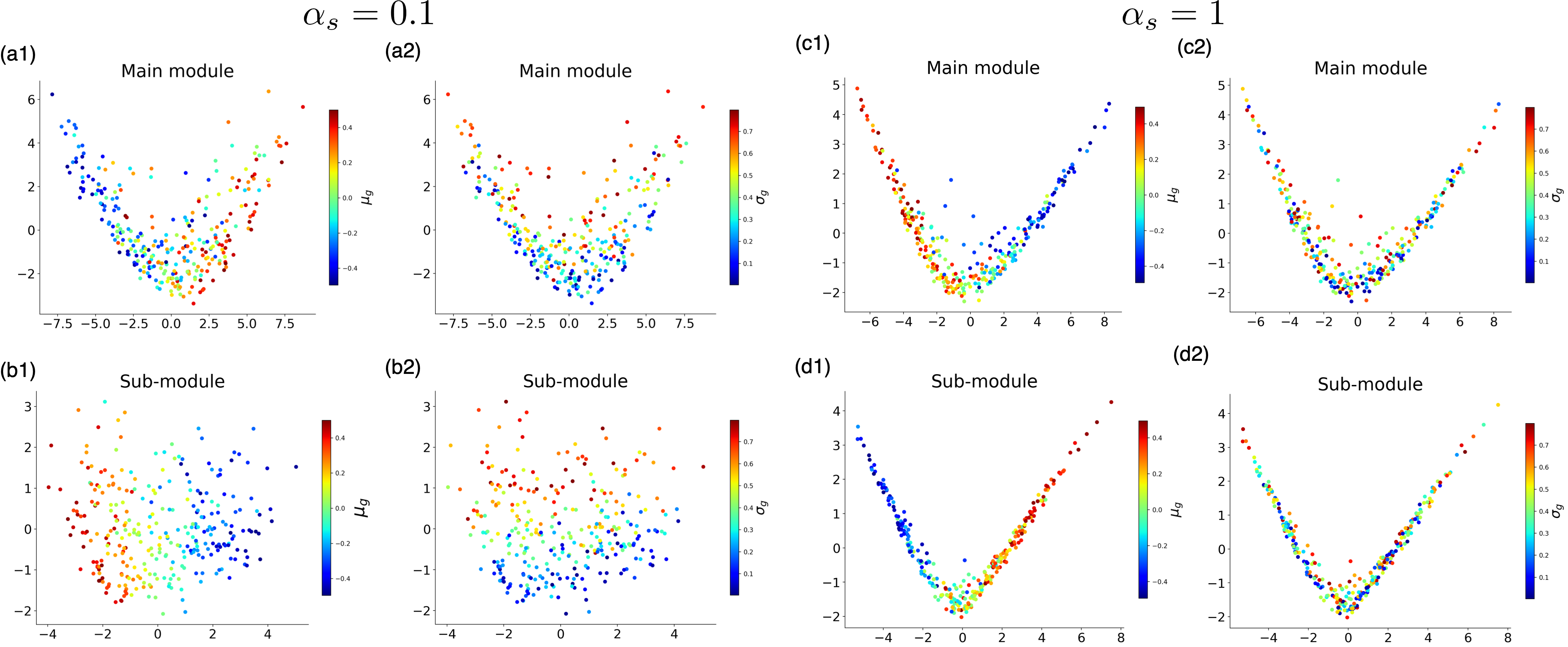}
  \caption{
    The neural activities of the main module ${\bf x}_m(\mu_g, \sigma_g)$ and sub-module ${\bf x}_s(\mu_g, \sigma_g)$ were plotted by the first and second principal component spaces. 
    (a,b) is the result of $\alpha=0.1$, and (c,d) is of $\alpha_s=1$. 
    (a1,c1) Main module, color-coded by $\mu_g$. 
    (a2, c2) Main module, color-coded by $\sigma_g$. 
    (b1,d1) Sub-module, color-coded by $\mu_g$. 
    (b2, d2) Sub-module, color-coded by $\sigma_g$. 
    300 data are plotted.
    }
  \label{fig:pca_internal_states}
\end{figure*}

Next, we investigated how the prior is represented by the main and sub-modules by visualizing the neural activity by principal component analysis(PCA)\citep{Mante2013,short-term-memory}.
First, ${\bf x}_m(\mu_g, \sigma_g)$ and ${\bf x}_s(\mu_g, \sigma_g)$ were computed for various $(\mu_g, \sigma_g)$ in a model with $\alpha_s =0.1$, and made PCA.
The results were projected on a plane using the first and second principal components and color-coded according to $\mu_g$ and $\sigma_g$ (Fig.\ref{fig:pca_internal_states}(a,b)).
The neural activity in the main module was loosely distributed on a one-dimensional manifold, represented by the first principal component(PC1). 
This PC1 approximately corresponded to the $\mu_g$ value, although the distinction was not clear.
In contrast, the activity in the sub-module was clearly represented by 2-dimensional manifolds, as in Fig.\ref{fig:pca_internal_states}(b2), where PC1 corresponds to $\mu_g$ and PC2 corresponds to $\sigma_g$, rather well.

Then, we performed the same analysis on the model with $\alpha_s=1$ (Fig.\ref{fig:pca_internal_states}(c,d)).
In this case, the manifolds of neural activities for the main and sub-modules did not change significantly. 
Both were represented in a one-dimensional manifold corresponding to $\mu_g$; there was no axis corresponding to $\sigma_g$. 
The decodability of $\sigma_g$ achieved in the internal states of sub-module with $\alpha_s=0.1$ was not observed for $\alpha_s=1$. 
In fact, the coefficient of determination when $\sigma_g$ was calculated by Ridge regression from the internal state of the sub-module with $\alpha_s=0.1$ was $0.68$, while that using the sub-module with $\alpha_s=1$ is $-0.03$.
This suggests that the model with $\alpha_s \sim 0.1$ can better distinguish the input's variance from noise to accurately perform Bayesian inference.

When the generator changed rapidly, the variance of the prior was larger than the variance of the generator, as shown in the SI for the case with $\alpha_s=0.1$.
When $\sigma_g$ was large, as seen from Eq.\ref{eq:y_opt}, the influence of the observed signal $s$ was larger than that of $\mu_g$, allowing the model to "keep up" with large changes in the observed signal. 
This explains the higher adjustability to rapid generator changes as seen in Fig.\ref{fig:adjustability}.

\subsection*{Effects of different time scales}
To examine the impact of $\alpha_s$ differences on Bayesian inference accuracy in detail, we considered how each model with $\alpha_s=1$ and $\alpha_s=0.1$ represents prior as a function of the input signal. 
As seen in Fig.\ref{fig:pca_internal_states}, when the generator is constant, the internal state of the neural network corresponds with the state of the generator $(\mu_g, \sigma_g)$. 
Conversely, when the generator changes, the internal state at a certain time does not necessarily correspond to the state of the generator at that time because some time is needed to estimate the state of the prior after the generator switches. 
Let $\mu_p$ and $\sigma_p$ be the mean and standard deviation of prior used by the neural network to compute $y(t)$.
Hence, $\mu_p$ must memorize the input $s(t)$ for a certain time in the form of

\begin{equation}
    \mu_p \simeq \sum a_k s(t-k).
    \label{eq:approximation_of_mu_p}
\end{equation}

To examine how many past steps $k$ are memorized, $\mu_p$ must be estimated.
This can be achieved by estimating $\mu_p$ from the internal state ${\bf x}(t)$.

First, we calculated the internal state ${\bf x}$ for the observed signal with a fixed generator instead of a time-varying case. 
Then, we found the transformation matrix $W_{\mu_p}$ from the internal state ${\bf x}$ to the recognized prior $\mu_p$ by assuming that $\mu_p$ can be represented by linear transformations of the internal state as $\mu_p \simeq W_{\mu_p}{\bf x}$.
This transformation matrix $W_{\mu_p}$ was obtained by a pseudo-inverse method\citep{reservoir_computing}(SI).

Next, we obtained ${\bf x}(t)$ against the time-varying signal with a probability of $p_t=0.03$. 
By applying the above transformation matrix, $W_{\mu_p}$ to ${\bf x}(t)$ obtained at this time, the prior $\mu_p$ was estimated accordingly.
The state of the prior was thus obtained for the time series of the observed signal $s(t)$. 

Then, $a_k$ in Eq.\ref{eq:approximation_of_mu_p} was obtained to minimize the difference between the two sides of Eq.\ref{eq:approximation_of_mu_p}. 
Because the obtained coefficients correspond to the contribution of the signal before $k$ time steps, we could estimate the extent to which the neural network uses past information when estimating the prior.

\begin{figure}
  \centering
  \includegraphics[width=12cm]{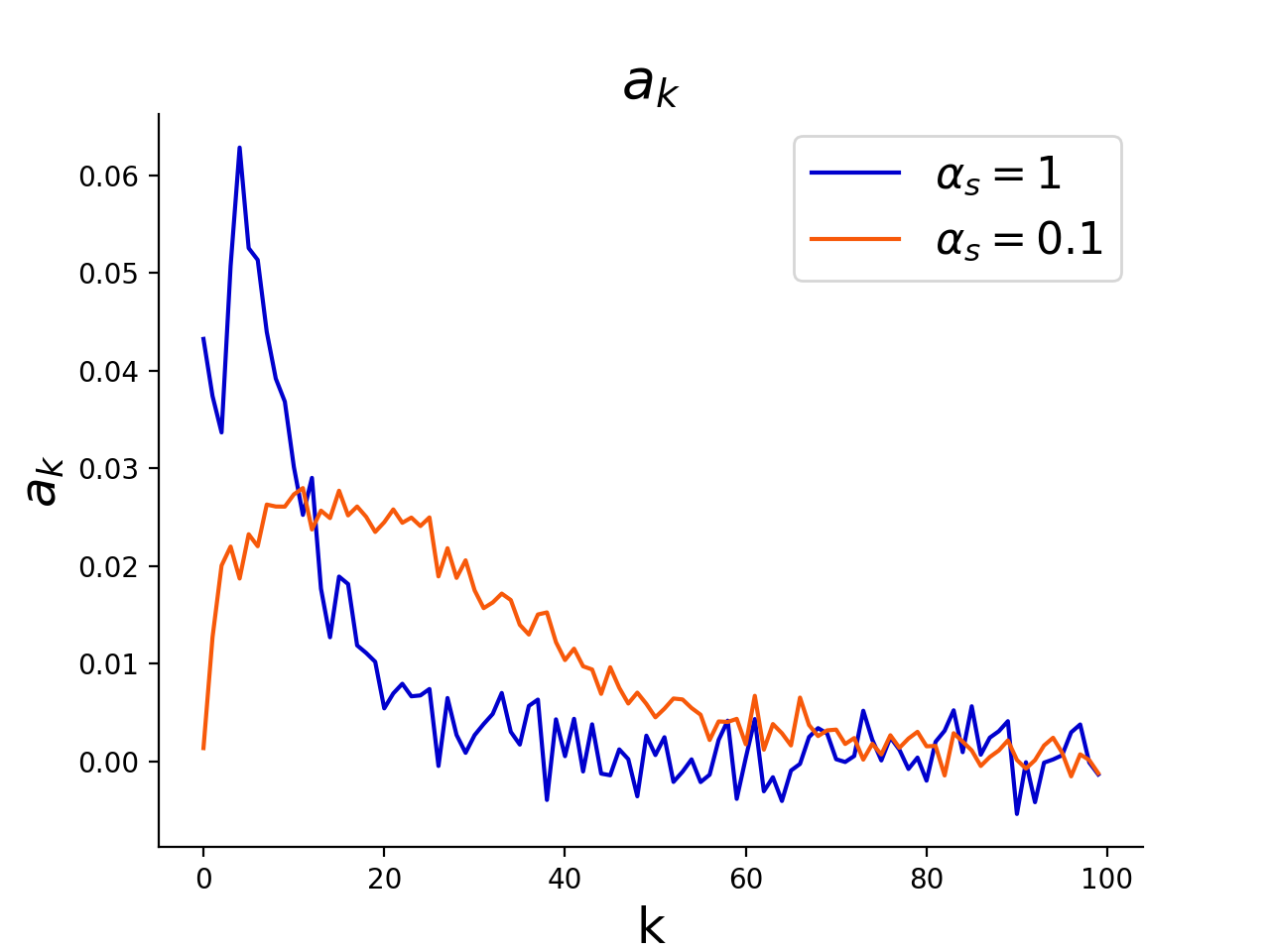}
  \caption{
    $a_k$ defined by Eq.\ref{eq:approximation_of_mu_p} is plotted against $t$, for the model with $\alpha_s=1$ and $\alpha_s=0.1$ using 3000 data points.
    }
  \label{fig:recognize_alpha}
\end{figure}

The estimated coefficients of Eq.\ref{eq:approximation_of_mu_p} were plotted against $k$(Fig.\ref{fig:recognize_alpha}), revealing that the model with $\alpha_s=0.1$ used more past information in estimating prior information than the model with $\alpha_s=1$.
This difference in time windows leads to a difference in accuracy for prior encoding.

\section*{Results2: Modular structure organization and time-scale separation by learning}
So far, we investigated neural networks with fixed and modular structures along fixed time scales and demonstrated that those with fast and slow modules effectively represented the prior distribution.
Then, we investigated whether such a structure would emerge by training a neural network to predict $y_{true}$.
We again used the same neural network model as the normal RNN.

\begin{equation}
    {\bf x}(t+1)=({\bf I}-{\boldsymbol \alpha}){\bf x}(t)+{\boldsymbol \alpha}{\rm ReLU}(W_{in}{\bf u}(t)+W_{rec}{\bf x}(t)) + \sqrt{\boldsymbol \alpha}{\boldsymbol \xi}, 
\end{equation}
where ${\boldsymbol \alpha}$ represents a vector of time scales of neurons consisting of $\alpha_i$.
These ${\boldsymbol \alpha}$ values, as well as elements of $W$, change by training to start from initial values set randomly according to $\mathcal{N}(0.5, 0.1)$. 
During training, each matrix $W$ and ${\boldsymbol \alpha}$ are optimized according to the gradient descent method\citep{neural_heterogeneity} at each step.
The number of neurons in the recurrent layer of the neural network was set to 80.

\begin{figure*}
  \centering
  \includegraphics[width=15cm]{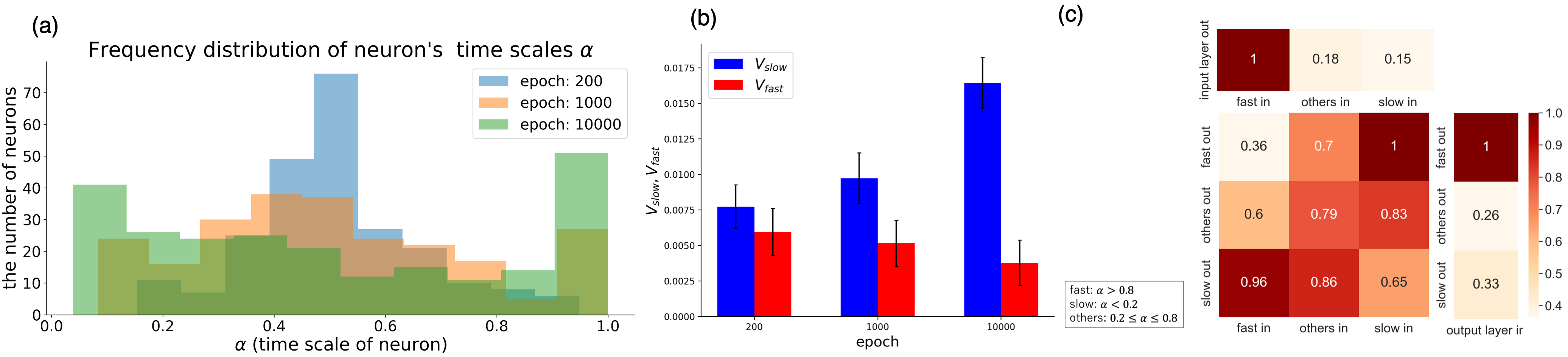}
  \caption{
  RNN features obtained by learning when ${\boldsymbol \alpha}$ is variable by learning. 
  (a) Frequency distribution of ${\boldsymbol \alpha}$ for all neurons at 200, 1000, and 10000 learning epochs. At 10000 epochs, the learning process was complete.
  (b) Division of roles $V_{slow}$ and $V_{fast}$. 
  See the "Representation of the prior" section for definitions of $V_{slow}$ and $V_{fast}$; $V_{slow}\ (V_{fast})$ was computed for neurons with $\alpha < 0.2\ (\alpha > 0.8)$ respectively. 
  (c) The average degree of RNN connections of 10000 epochs. Connections between input layer, recurrent neurons with $\alpha < 0.2$, $0.2\leq \alpha \leq 0.8$, $\alpha > 0.8$, and output layer. 
  Each was normalized so that the maximum value was 1.
  }
  \label{fig:trainable_alpha}
\end{figure*}

The change in ${\boldsymbol \alpha}$ distribution during the learning task is shown in Fig.\ref{fig:trainable_alpha}(a).
As shown, ${\boldsymbol \alpha}$ split into two groups over the learning period: one with large values close to 1 and the other with small values near 0.1.

Next, we measured the contribution of prior representation as examined in the "Representation of the prior" section for groups of neurons with large values ${\boldsymbol \alpha}$ (neurons with $\alpha_i>0.8$) and groups of neurons with small values ${\boldsymbol \alpha}$(neurons with $\alpha_i<0.2$) for three epochs in the learning process(Fig.\ref{fig:trainable_alpha}(b)).
We found that after 10000 epochs, the slow neurons were responsible for the representation of prior distribution, as in the model with $\alpha_s=0.1$ in the fixed time scale setting.

Finally, we investigated the neural network structure shaped by training. 
In Fig.\ref{fig:trainable_alpha}(a), the recurrent layer neurons of the network of epoch 10000 was split into the three groups, divided by the magnitude of $\alpha_i$, slow neurons with $\alpha_i<0.2$, fast neurons with $\alpha_i > 0.8$, and $0.2\leq \alpha_i \leq 0.8$ neurons as the others. 
The average connectivity between the input layer, each group, and the output layer is shown in Fig.\ref{fig:trainable_alpha}(c)\citep{task-representation}.
The connection from the input layer to the group of fast neurons and that from the fast neurons to the output layer were distinctively larger than those to or from the slow neurons.
Among connections within the recurrent layer, those between the fast and slow neurons were larger than others.
In summary, a modular structure, shown in Fig.\ref{fig:hierarchical_rnn}(b), emerged through learning alone.

\section*{Discussion}
In this study, we demonstrated that neural networks with slow and fast activity modules play an essential role in the prior representation for Bayesian inference.
We set up a task to predict a time-varying signal under noise that could be estimated by Bayesian inference and trained RNNs with or without modular structure and with or without time scale differences.

The RNN could learn to approximate Bayesian inference using prior(approximating the generator distribution) in all conditions tested.
However, the accuracy was higher in the modular RNN; further, the accuracy was significantly higher when the time scale of the sub-module was moderately slower than that of the main module.
In addition, the increase in accuracy was pronounced against a rapidly varying input, for which it was necessary to generate a prior that changes quickly.
To achieve such accuracy with a slow sub-module, the sub-module was found to dominantly represent the prior, indicating role differentiation between representation of the prior and representation of the observed signal (likelihood).
Of note, such functional differentiation is caused by differences in time scales.
This result is consistent with experimental observations in the brain in which areas that code the prior and likelihood in Bayesian inference are different\citep{differential_representation, Chan7817, dAcremont10887}.
Finally, it was shown that a modular structure with distinct time scales was spontaneously organized in the RNN by learning. 

It is important to note that a relatively slow time scale of the neuron population encoding the prior is required, but the difference between fast and slow neurons should not be excessive.
If the time scale is too small, the accuracy is decreased (Fig.\ref{fig:bayesian_optimality}) in which case the sub-module is not responsible for representing the prior (Fig.\ref{fig:differentiation_of_prior_representation}).
This is because prior construction requires a larger time span to address changes in external input for a neural network with such a slow time scale.
Therefore, we suggest that there is an optimal time scale for the slow sub-module.
Future research should investigate how this optimal time scale depends on the time scale of environmental changes.

It has been suggested that the time scale of neurons slows down hierarchically from the area where the signal is directly applied to the area where information is proceed\citep{hierarchy_of_intrinsic_timescale, diversity_intrinsic_timescale, brain_and_its_time}.
This time scale hierarchy with a modular structure\citep{hierarchical_task} is suggested to be relevant to information processing\citep{multi-timescales, hierarchical_task, multi_scale_dynamics}.
Our study showed that modular structures with two-level time scales could deal with slowly changing inputs.
A deeper modular structure with multiple time scales may be necessary to deal with further complex changes in environments. 
With such a structure, Bayesian inference against complex temporal changes could be achieved by extrapolating the results of this study.
Further research verifying this finding will elucidate the significance of hierarchical structuring in the brain. 
It is noteworthy that the time scale separation was not only found to be influential for accurate Bayesian inference but also emerged from learning in our simulation. 
Considering these findings, a similar process may be expected in evolution\citep{doi:10.1063/5.0019116}.

The modular network with slow/fast time scales could integrate out noise and distinguish the average change in the inputs from fast noise.
In fact, the network could effectively predict temporal changes in the input, even under rapidly changing conditions.
The brain must adapt to time-varying, noisy inputs; hence, the performance of Bayesian inference by the network design reported herein is considered relevant to brain information processing.

We adopted a simple RNN and trained it using backpropagation.
red{Backpropagation} is often believed to be different from the learning algorithm implemented in the brain\citep{biologically_plausible1, biologically_plausible2}, so care should be taken when generalizing our results.
However, previous studies have also suggested that neural networks obtained by backpropagation can show similar behavior to that of the actual brain\citep{deeplearning-neuroscience1, deeplearning-neuroscience2, Mante2013, BARAK2013214, j.2018emergence, goal_driven_deep_learning, HAESEMEYER20191123}. 
With these considerations, our findings are considered to be relevant to the brain’s learning processes despite the potential limitation of backpropagation.

Unraveling the relationship between the structure of neural networks, neural dynamics, and the information processing performed by the brain is a primary goal in computational neuroscience\citep{MASTROGIUSEPPE2018609, role_of_population_structure, Computation_Through_Neural_Population_Dynamics, Shaping_dynamics}. 
In this study, the relevance of modular structure and time scale difference in neural dynamics to the representation of the prior in Bayesian inference is demonstrated, as well as their formation by learning\citep{LORENZ2011129, spontaneous_evolution_of_modularity}, which will support ongoing research in the field.

\subsection*{Data Availability}
Source codes for these models can be found at \href{https://github.com/tripdancer0916/slow-reservoir}{https://github.com/tripdancer0916/slow-reservoir}

\section*{Appendix}
\subsection*{Appendix A: Adjustment to rapid environment changes}
The trajectories of the internal state ${\bf x}_{sub}(t)$ are plotted by the first and second principal components in Fig. 6 for the cases in which generators A and B switch every 2 time steps and every 30 time steps. 
Generators A and B both have $\sigma_g=0.04$. 
In the case of switching every 30 time steps, they were located in the region taken by the internal state when $\sigma_g$ was small. In the case of switching every 2 time steps, they were located in the region taken by the internal state when $\sigma_g$ was large (See Fig. S1). 
This occurred because the generators switched so rapidly that the RNN recognized that the signal was created by a generator with a large variance. 
This made it possible to switch $y(t)$ quickly because the information of the observed signal $s$ was prioritized over the prior information when calculating the output $y$.

\begin{figure}[H]
\centering
\includegraphics[width=10cm]{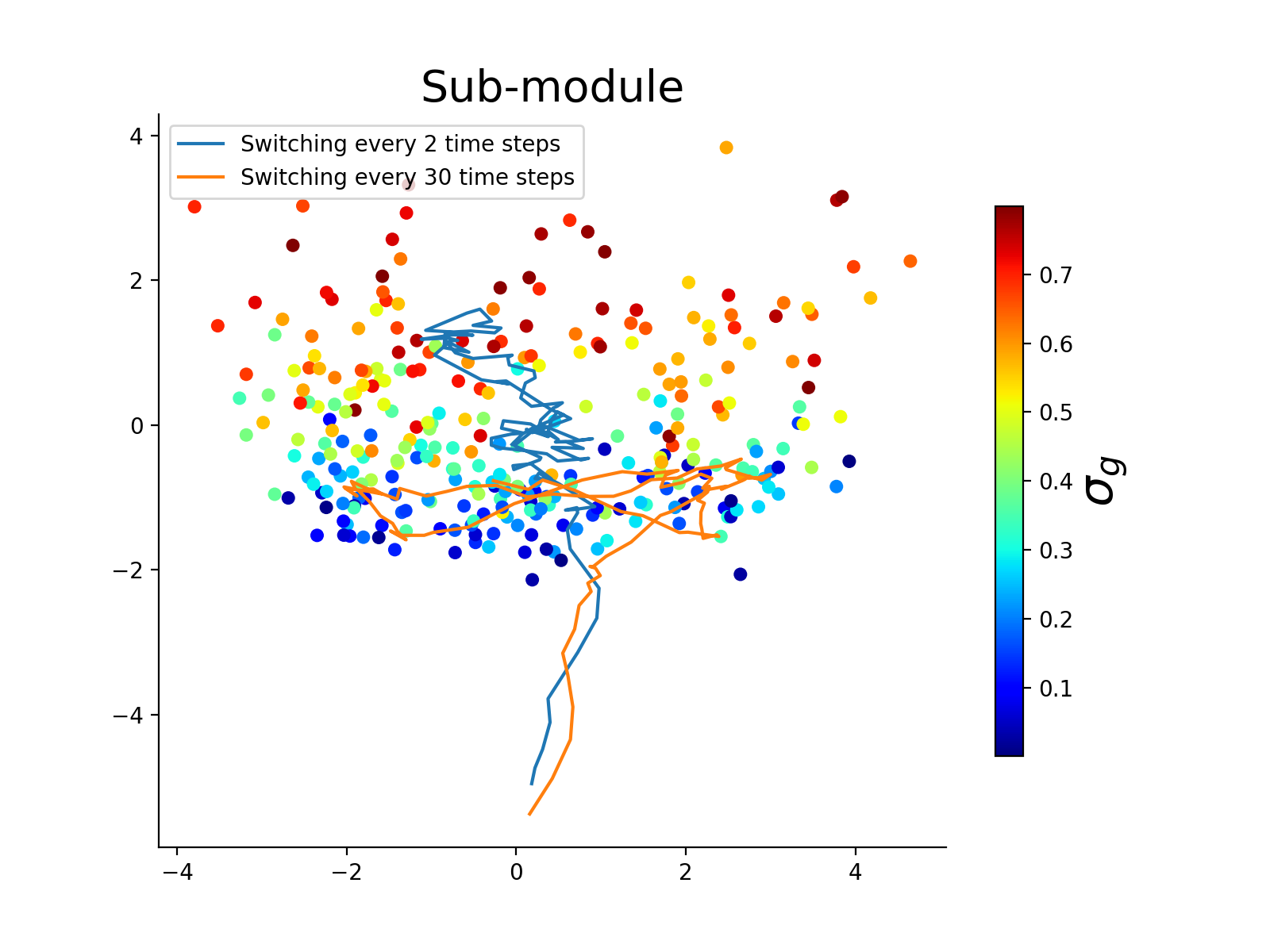}
\label{fig:adjust}
\caption{Trajectory of the internal state ${\bf x}_{sub}(t)$ of the sub-module when generator A($(\mu_A, \sigma_A^2)=(-0.5, 0.04)$) and generator B($(\mu_A, \sigma_A^2)=(0.5, 0.04)$) switch alternately.}
\end{figure}

\subsection*{Appendix B: Results of the RNN with $\alpha_m=\alpha_s=0.1$}
We argue that the slower time scale of the sub-module relative to the main module is important for accurate Bayesian inference. 
Here, to investigate whether the difference in time scale or the slower time scale itself were more influential, we trained an RNN with $\alpha_s=\alpha_m=0.1$ and examined its accuracy.
We found that when $(\alpha_m, \alpha_s)=(0.1,0.1)$, the MSE was larger, and the accuracy was worse than that in the cases with $(\alpha_m, \alpha_s)=(1,1)$ and $(\alpha_m, \alpha_s)=(1,0.1)$, as shown in Fig. S2. 
Therefore, it is not simply the slower time scale of the neurons but the time scale difference between the main and sub-modules that facilitate accurate Bayesian inference.

\begin{figure}
\centering
\includegraphics[width=10cm]{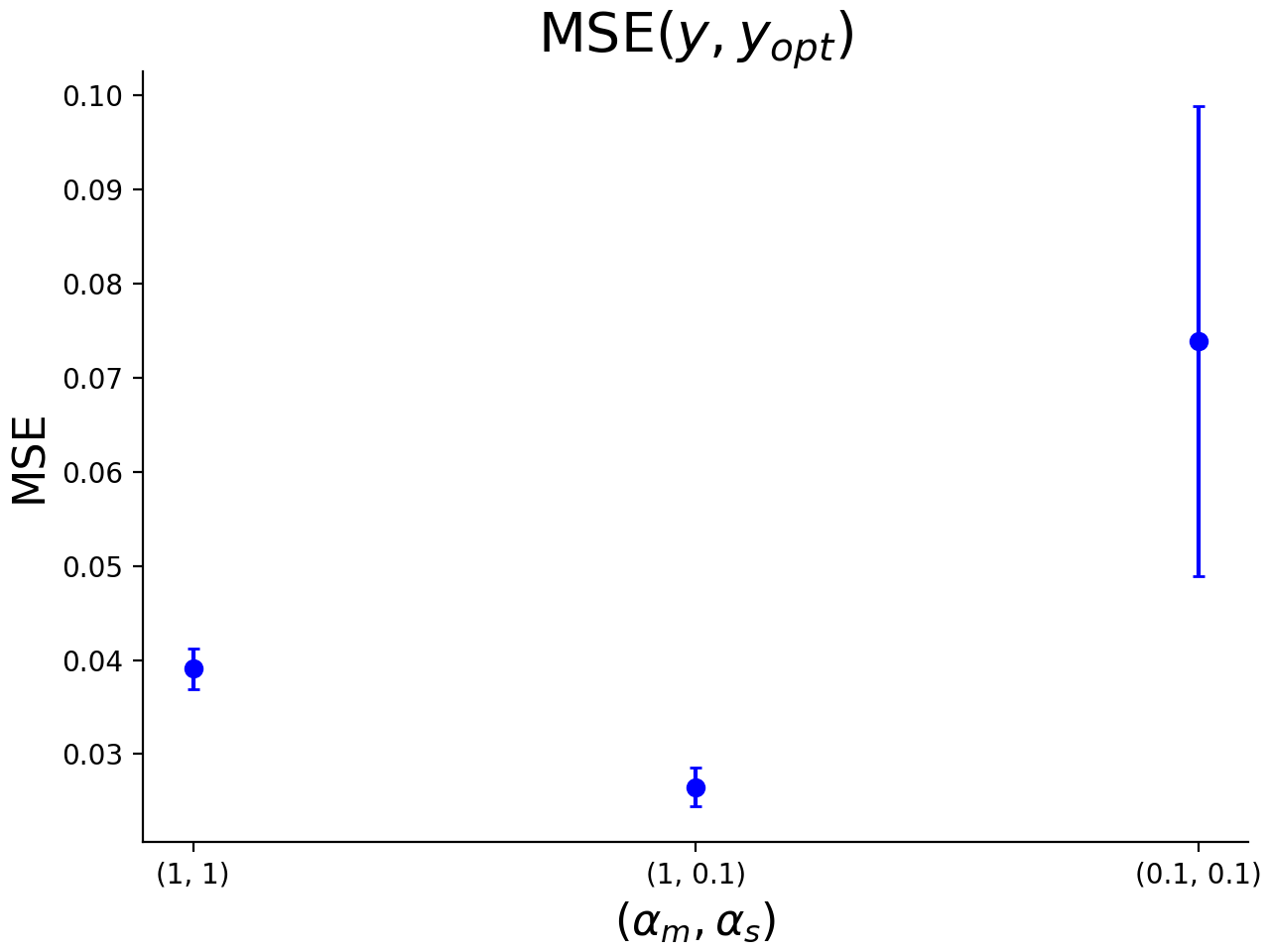}
\label{fig:comparison}
\caption{Mean squared error between the optimal value $y_{opt}(t)$ and the output of RNN $y(t)$, plotted against the setting $(\alpha_m, \alpha_s)=(1,1),(1,0.1),(0.1,0.1)$.}
\end{figure}

\subsection*{Appendix C: Detailed analysis of the time scale difference effect}
To obtain the transformation matrix $W_{\mu_p}$, we created a data vector $M_g$ that arranges $\mu_g$ and a data matrix $X$ that arranges the internal state $\bf x$ obtained when fixed $\mu_g$ was input as follows:

\begin{equation*}
    M_g = (\mu_g^1, \mu_g^2, ...)^{\mathsf T}, \ X = ({\bf x}^1, {\bf x}^2, ...)
\end{equation*}
Then, we attempted to find matrix $W_{\mu_p}$ such that $M_g \simeq W_{\mu_p}X$.
Using the Moore-Penrose pseudo inverse, we can find the best-fit matrix as $W_{\mu_p}=M_g X^\dag$\citep{penrose_1955}.
Let $\mu_p$ be the result of the transformation by $W_{\mu_p}$. 
As Fig. S3(a) shows, $\mu_g\simeq \mu_p$ is valid.
To find $a_k$, we calculated ${\bf x}(t)$ in the case that $\mu_g$ varies randomly using $W_{\mu_p}$ and obtained $\mu_p$. 
Then, we created a data vector $M_p$ that arranges $\mu_p$ and a data matrix $S$ that arranges ${\bf s}=(s(t-1), s(t-2), ..., s(t-K))^{\mathsf T}$.
Using the Moore-Penrose pseudo inverse, we found ${\bf a}=(a_1,a_2,...,a_K)^{\mathsf T}$.
As Fig. S3(a) shows, $\mu_p \simeq \sum_k^K a_k s(t-k)$ was valid.

\begin{figure}[H]
\centering
\includegraphics[width=14cm]{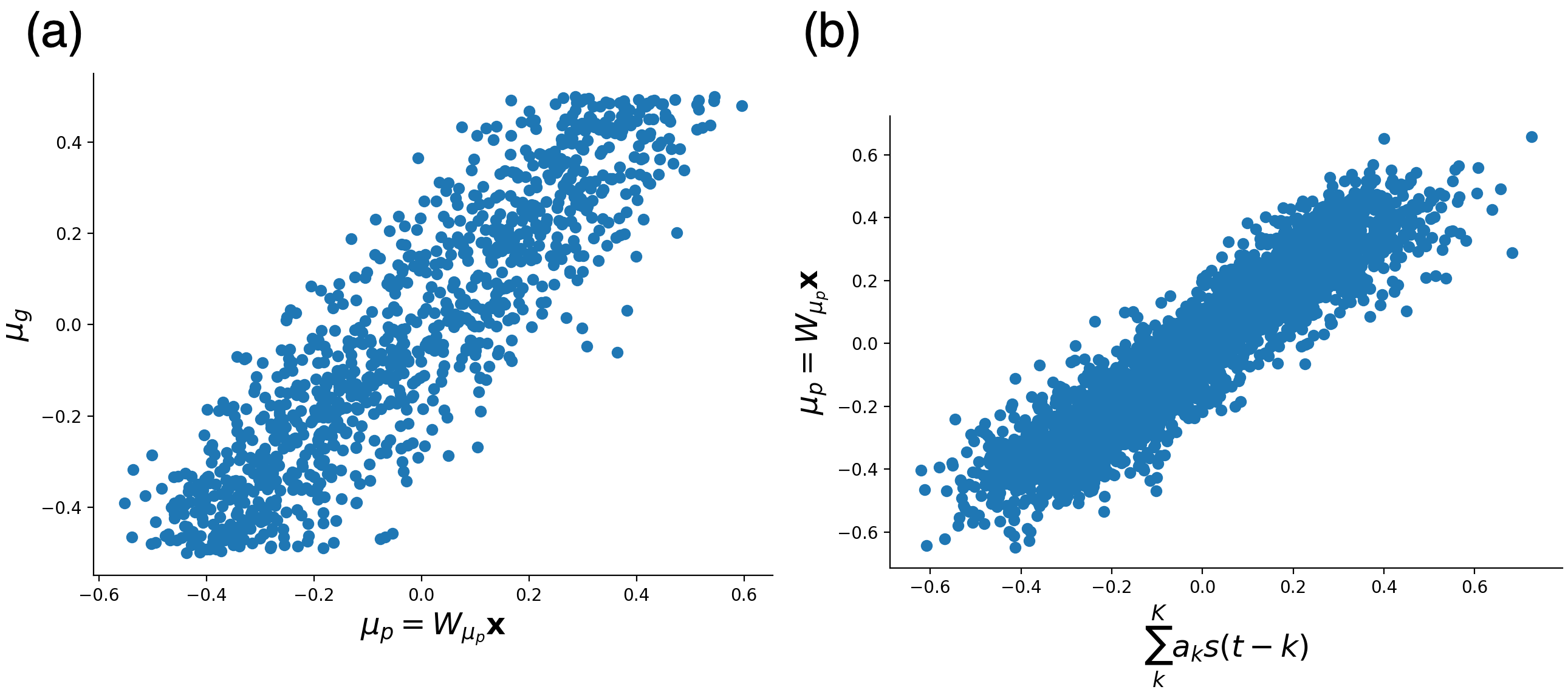}
\label{fig:decoding}
\caption{(a)Comparison between the estimated mean of prior $\mu_p$ and the mean of generator $\mu_g$.
(b)Comparison between the linear weighted sum of past signals $s(t-k)$ and the estimated mean of prior $\mu_p$.}
\end{figure}

\bibliographystyle{unsrtnat}
\bibliography{references}  %%% Uncomment this line and comment out the ``thebibliography'' section below to use the external .bib file (using bibtex) .





% \end{thebibliography}

\end{document}